**ORIGINAL PAPER**

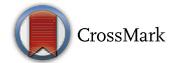

# Heat capacity and thermal expansion of metal crystalline materials based on dynamic thermal vibration

**Jieqiong Zhang[1] · Junzhi Cui[1,2] · Zihao Yang[1] · Yifan Yu[2]**

**Abstract**

A novel approach based on dynamic thermal vibration is proposed to calculate the heat capacity and thermal expansion coefficient (TEC) for metal crystalline materials from 0 K to the melting point. The motion of metal atomic clusters is decomposed into structural deformation and thermal vibration. Then thermal vibration equations are established by the fourth-order Taylor expansion of Hamilton at the transient structural deformation position $\bar{\mathbf{x}}$. As a result, the thermal vibration frequencies dynamically change with the structural deformation positions and temperatures. A parameter $\bar{\delta}(\bar{\mathbf{x}}, T)$ is newly introduced to illustrate how the thermal vibration frequencies vary with the temperature $T$. Besides, the modified temperature-dependent Grüneisen parameter $\bar{\gamma}(\bar{\mathbf{x}}, T)$ is given. Finally, the formulae of heat capacity and TEC for metal crystalline materials are derived from the dynamic thermal vibration frequencies and $\bar{\delta}(\bar{\mathbf{x}}, T)$ as well as $\bar{\gamma}(\bar{\mathbf{x}}, T)$. The numerical results of heat capacity and TEC for metals Cu, Al, Au, Ag, Ni, Pd, Pt and Pb show a temperature dependence and agree well with the experimental data from 0 K to the melting point. This work suggests an efficient approach to calculate thermodynamic properties of metal materials for a wide range of temperatures, up to the melting point.

**Keywords** Metal crystalline materials · Heat capacity · Thermal expansion coefficient · Dynamic thermal vibration · Temperature dependence

## 1 Introduction

Thermodynamic properties of metal crystalline materials are related to the thermal vibration behaviors of atomic clusters. With harmonic approximation [1–3], the Hamiltonian is obtained by the second-order Taylor approximation at the lattice points. And the thermal vibration of atomic clusters is reduced to a set of independent oscillators vibrating around the lattice points at 0 K. The frequencies of these harmonic oscillators are independent of the volume and temperature, so that the heat capacity becomes constant at elevated temperatures and the heat capacity at constant pressure is equal

✉ Jieqiong Zhang
jieqiongzhang@mail.nwpu.edu.cn

✉ Zihao Yang
yangzihao@nwpu.edu.cn

1  Department of Applied Mathematics, Northwestern Polytechnical University, Xi'an 710072, China

2  LSEC, ICMSEC, Academy of Mathematics and Systems Science, Chinese Academy Sciences, Beijing 100090, China

to the heat capacity at constant volume. In addition, there is no thermal expansion [2–6].

Quasi-harmonic approximation (QHA) is further developed by assuming that the thermal vibration frequency for the $i$th oscillator depends on the volume $V$, namely $\omega_i = \omega_i(V)$ [4,5,7]. The Grüneisen parameter, denoted by $\gamma_i = -\partial \ln \omega_i / \partial \ln V$, is introduced, which plays a vital role in expressing heat capacity and thermal expansion coefficient (TEC). With QHA, Motuzzi et al. [8] studied the heat capacities and TECs of 14 nonmagnetic cubic metals at low temperatures. They provided a quasi-harmonic treatment by applying the harmonic approximation at each temperature stage and Grüneisen parameter was a constant. The heat capacity and thermal expansion could be simulated well with QHA at low temperatures. However, QHA is inaccurate at high temperatures [4,5,7]. There have been some researches focusing on improving QHA. Gan [9] derived a compact matrix expression for linear TECs of metal chalcogenides based on mode-dependent Grüneisen parameters and the Grüneisen parameter was regarded as a function of thermal vibration frequencies. Besides, the effects of the higher-order terms in the Taylor expansion of Hamiltonian were investigated by Glensk et al. [10–14]. They studied the interactions between oscillators caused by the higher-order terms and derived the heat capacity and TEC from the modified free energy with first principle theory. In addition, other researchers [15–19] focused on the thermal vibration frequencies by combining the ab initio density functional theory (Ab-DFT) and density functional perturbation theory with the QHA and Grüneisen parameter. The interactions of the electrons are considered by taking the local density approximation (LDA) or generalized gradient approximation (GGA) as the exchange-correlation energy. The temperature-dependent heat capacity and TEC are computed and their results show smaller errors to the experiment data by comparing with the harmonic quantities.

Different from the previous works, the thermal vibration dynamically influenced by the temperature and stress is analyzed in this paper, which is essentially based on the thermo-mechanical coupling mechanism. In the atomistic-continuum coupled (ACC) model proposed by Cui et al. [20–22], the Hamiltonian is evaluated by a second-order Taylor expansion at the transient structural deformation positions and the thermal vibration dynamically changes with the state of structural deformation. They provided the formula of heat capacity for crystalline metals Cu at room temperature. Besides, Albert et al. [23] studied the thermodynamic properties by applying a third-order Taylor approximation to Hamiltonian based on the quasi-harmonic approximation. The heat capacity, TEC and elastic constant were calculated at finite temperatures.

In this work, the heat capacity and TEC of metal crystalline materials are investigated, based on the ACC model. The Hamiltonian is obtained approximately by the fourth-order Taylor expansion at the transient structural deformation position $\bar{\mathbf{x}}$. It should be noted that the transient structural deformation position is quite different from the lattice point at $0\,\mathrm{K}$ under the harmonic or quasi-harmonic approximation. The transient structural deformation position changes with the temperature and the stress field. Therefore, the thermal vibration frequency of the $i$th oscillator dynamically depends on the transient structural deformation position $\bar{\mathbf{x}}$ and the temperature, namely $\omega_i^d = \omega_i^d(\bar{\mathbf{x}}, T)$. A parameter $\delta_i(\bar{\mathbf{x}}, T) = -\partial \ln \omega_i^d / \partial \ln T$ is introduced to describe the change rate of the thermal vibration frequency with respect to the temperature. In addition, the Grüneisen parameter $\bar{\gamma}(\bar{\mathbf{x}}, T)$ is modified as a temperature-dependent parameter. Both two parameters are of great importance in the calculation of heat capacity and TEC, especially at high temperatures.

The remainder of this paper is outlined as follows. In Sect. 2, we establish the thermal vibration equations by the fourth-order Taylor expansion of Hamiltonian at transient structural deformation positions and obtain the dynamic thermal vibration frequencies by solving the equations with Jacobi elliptic expansion method. Then thermodynamic properties are derived from dynamic thermal vibration frequencies in Sect. 3. Sections 4 and 5 display the algorithm procedure and the numerical examples of heat capacity and TEC for Cu, Al, Au, Ag, Ni, Pd and Pt, respectively. Finally, we draw the conclusions in Sect. 6.

## 2 Analysis of dynamic thermal vibration

In this section, the motion of atomic system influenced by the temperature and stress is decomposed into structure deformation and thermal vibration. The thermal vibration equations are established in transient equilibrium state, i.e. the equilibrium state at different structural deformations and temperatures. By solving the equations with Jacobi elliptic-function method, the thermal vibration frequencies that dynamically change with the structural deformation and temperature are obtained.

### 2.1 The decomposition of atomic motion

A atomic cluster consisting of $N$ atoms in a domain $\Omega_R$ is called as the representative volume element (RVE) and it involves several crystalline lattices to characterize the microstructure of crystalline materials. Considering the interactions between the atoms in $\Omega_R$ and their neighboring atoms, an extended representative volume element (ERVE) $\Omega_E$ with $N_E$ atoms is defined [21,22], which contains all the atoms in $\Omega_R$ and their neighboring atoms. The RVE and ERVE are shown in Fig. 1. The motion of the $i$th atom $\Omega_R$ can be decomposed into two parts: structural deformation and thermal vibration (high-frequency vibration) [21,22]

$$\begin{aligned} \mathbf{x}_i &= \bar{\mathbf{x}}_i + \tilde{\mathbf{x}}_i, \\ \dot{\mathbf{x}}_i &= \dot{\bar{\mathbf{x}}}_i + \dot{\tilde{\mathbf{x}}}_i, \end{aligned} \tag{1}$$

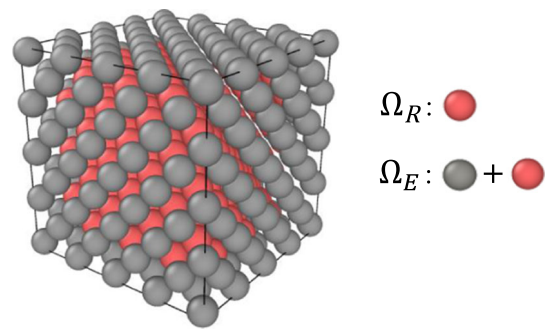

**Fig. 1** The RVE $\Omega_R$ and ERVE $\Omega_E$ (Color online)

where $\mathbf{x}_i = (x_i^1, x_i^2, x_i^3)$ is the instantaneous position of the $i$th atom with Cartesian coordinate; $\dot{\mathbf{x}}_i = (\dot{x}_i^1, \dot{x}_i^2, \dot{x}_i^3)$ is the instantaneous velocity of the $i$th atom; $\bar{\mathbf{x}}_i = (\bar{x}_i^1, \bar{x}_i^2, \bar{x}_i^3)$ and $\dot{\bar{\mathbf{x}}}_i = (\dot{\bar{x}}_i^1, \dot{\bar{x}}_i^2, \dot{\bar{x}}_i^3)$ denote the position and velocity of structural deformation for the $i$th atom, respectively; $\tilde{\mathbf{x}}_i = (\tilde{x}_i^1, \tilde{x}_i^2, \tilde{x}_i^3)$ and $\dot{\tilde{\mathbf{x}}}_i = (\dot{\tilde{x}}_i^1, \dot{\tilde{x}}_i^2, \dot{\tilde{x}}_i^3)$ denote the displacement and velocity of thermal vibration for the $i$th atom, respectively. The instantaneous position vector for all atoms is defined as $\mathbf{x} = (\mathbf{x}_1, \mathbf{x}_2, \ldots, \mathbf{x}_N) = (x_1^1, x_1^2, x_1^3, \ldots, x_N^1, x_N^2, x_N^3)$ and the velocity vector $\dot{\mathbf{x}}$ is defined similarly.

The structural deformation position, $\bar{\mathbf{x}} = \bar{\mathbf{x}}(f, T)$, is influenced by the stress $f$ and the temperature $T$. In this paper, we only consider the structural deformation affected by temperature, namely $f = 0$. Besides, the RVE considered is in equilibrium state at different structural deformations and temperatures.

## 2.2 The fourth-order Taylor expansion of Hamiltonian

The Hamiltonian of RVE is expressed as

$$H(\bar{\mathbf{x}} + \tilde{\mathbf{x}}) = K(\dot{\bar{\mathbf{x}}} + \dot{\tilde{\mathbf{x}}}) + U(\bar{\mathbf{x}} + \tilde{\mathbf{x}}), \tag{2}$$

where $K(\dot{\bar{\mathbf{x}}} + \dot{\tilde{\mathbf{x}}})$ is the kinetic energy and $U(\bar{\mathbf{x}} + \tilde{\mathbf{x}})$ is the total potential energy of RVE. And $K(\dot{\bar{\mathbf{x}}} + \dot{\tilde{\mathbf{x}}})$ is shown as follows

$$K(\dot{\bar{\mathbf{x}}} + \dot{\tilde{\mathbf{x}}}) = \frac{1}{2}(\dot{\bar{\mathbf{x}}} + \dot{\tilde{\mathbf{x}}})' \mathbf{M} (\dot{\bar{\mathbf{x}}} + \dot{\tilde{\mathbf{x}}}), \tag{3}$$

where $\mathbf{M} = \mathrm{diag}(m_1, m_1, m_1, m_2, m_2, m_2, \ldots, m_N, m_N, m_N)$ is the mass matrix; $m_i$ is mass of the $i$th atom; the superscript $'$ denotes the transpose of vector or matrix. The total potential energy $U(\bar{\mathbf{x}} + \tilde{\mathbf{x}})$ is given by

$$U(\bar{\mathbf{x}} + \tilde{\mathbf{x}}) = \frac{1}{2} \sum_{i=1}^{N} \sum_{j=1, j \neq i}^{N_E} U_i(\bar{\mathbf{x}}_i + \tilde{\mathbf{x}}_i, \bar{\mathbf{x}}_j + \tilde{\mathbf{x}}_j), \tag{4}$$

where $U_i(\bar{\mathbf{x}}_i + \tilde{\mathbf{x}}_i, \bar{\mathbf{x}}_j + \tilde{\mathbf{x}}_j)$ is the potential energy of the $i$th atom influenced by the $j$th atom, $N$ and $N_E$ are the number of atoms in RVE $\Omega_R$ and ERVE $\Omega_E$, respectively.

Since the thermal vibration displacement is a small quantity, the total potential energy can be approximately obtained by the fourth-order Taylor expansion at the structural deformation position $\bar{\mathbf{x}}$ as follows

$$\begin{aligned}
U(\bar{\mathbf{x}} + \tilde{\mathbf{x}}) = {}& U_0(\bar{\mathbf{x}}) + \sum_{i=1}^{N} \sum_{\alpha=1}^{3} \left. \frac{\partial U(\mathbf{x})}{\partial x_i^\alpha} \right|_{\mathbf{x}=\bar{\mathbf{x}}} \cdot \tilde{x}_i^\alpha \\
& + \frac{1}{2!} \sum_{i,j=1}^{N} \sum_{\alpha,\beta=1}^{3} \left. \frac{\partial^2 U(\mathbf{x})}{\partial x_i^\alpha \partial x_j^\beta} \right|_{\mathbf{x}=\bar{\mathbf{x}}} \cdot \tilde{x}_i^\alpha \tilde{x}_j^\beta \\
& + \frac{1}{3!} \sum_{i,j,k=1}^{N} \sum_{\alpha,\beta,\gamma=1}^{3} \left. \frac{\partial^3 U(\mathbf{x})}{\partial x_i^\alpha \partial x_j^\beta \partial x_k^\gamma} \right|_{\mathbf{x}=\bar{\mathbf{x}}} \cdot \tilde{x}_i^\alpha \tilde{x}_j^\beta \tilde{x}_k^\gamma \\
& + \frac{1}{4!} \sum_{i,j,k,h=1}^{N} \sum_{\alpha,\beta,\gamma,\tau=1}^{3} \left. \frac{\partial^4 U(\mathbf{x})}{\partial x_i^\alpha \partial x_j^\beta \partial x_k^\gamma \partial x_h^\tau} \right|_{\mathbf{x}=\bar{\mathbf{x}}} \\
& \cdot \tilde{x}_i^\alpha \tilde{x}_j^\beta \tilde{x}_k^\gamma \tilde{x}_h^\tau,
\end{aligned} \tag{5}$$

where $U_0(\bar{\mathbf{x}})$ is the potential energy at the structural deformation position and is expressed as

$$U_0(\bar{\mathbf{x}}) = \frac{1}{2} \sum_{i=1}^{N} \sum_{j=1, j \neq i}^{N_E} U_i(\bar{\mathbf{x}}_i, \bar{\mathbf{x}}_j). \tag{6}$$

The following first-order derivative of potential energy $U(\mathbf{x})$ with respect to $x_i^\alpha$ is regarded as the force for the $i$th atom at $\alpha$ direction

$$\begin{aligned}
\left. \frac{\partial U(\mathbf{x})}{\partial x_i^\alpha} \right|_{\mathbf{x}=\bar{\mathbf{x}}} = {}& \frac{1}{2} \sum_{p=1, p \neq i}^{N_E} \left. \frac{\partial U_i(\mathbf{x}_i, \mathbf{x}_p)}{\partial x_i^\alpha} \right|_{\mathbf{x}=\bar{\mathbf{x}}} \\
& + \frac{1}{2} \sum_{l=1, l \neq i}^{N} \left. \frac{\partial U_l(\mathbf{x}_l, \mathbf{x}_i)}{\partial x_i^\alpha} \right|_{\mathbf{x}=\bar{\mathbf{x}}}.
\end{aligned} \tag{7}$$

For the second-order expansion term, a $3N \times 3N$ matrix $\mathbf{B}(\bar{\mathbf{x}}) = \left( \mathbf{b}_{ij}(\bar{\mathbf{x}}) \right)$ $(i, j = 1, 2, \ldots, N)$ is introduced as follows

$$\mathbf{B}(\bar{\mathbf{x}}) = \begin{pmatrix} \mathbf{b}_{11}(\bar{\mathbf{x}}) & \ldots & \mathbf{b}_{1N}(\bar{\mathbf{x}}) \\ \vdots & \ddots & \vdots \\ \mathbf{b}_{N1}(\bar{\mathbf{x}}) & \ldots & \mathbf{b}_{NN}(\bar{\mathbf{x}}) \end{pmatrix}, \tag{8}$$

where $\mathbf{b}_{ij}(\bar{\mathbf{x}}) = \left( b_{ij}^{\alpha\beta}(\bar{\mathbf{x}}) \right)$ is a $3 \times 3$ matrix and its components $b_{ij}^{\alpha\beta}(\bar{\mathbf{x}})$ $(\alpha, \beta = 1, 2, 3)$ are defined by the following two cases:

(i) $i = j$,

$$
\begin{aligned}
b_{ii}^{\alpha\beta}(\bar{\mathbf{x}}) &= \left.\frac{\partial^2 U(\mathbf{x})}{\partial x_i^\alpha \partial x_i^\beta}\right|_{\mathbf{x}=\bar{\mathbf{x}}} \\
&= \frac{1}{2} \sum_{p=1, p\neq i}^{N_E} \left.\frac{\partial^2 U_i(\mathbf{x}_i, \mathbf{x}_p)}{\partial x_i^\alpha \partial x_i^\beta}\right|_{\mathbf{x}=\bar{\mathbf{x}}} \\
&\quad + \frac{1}{2} \sum_{l=1, l\neq i}^{N} \left.\frac{\partial^2 U_l(\mathbf{x}_l, \mathbf{x}_i)}{\partial x_i^\alpha \partial x_i^\beta}\right|_{\mathbf{x}=\bar{\mathbf{x}}},
\end{aligned}
\tag{9}
$$

(ii) $i \neq j$,

$$
\begin{aligned}
b_{ij}^{\alpha\beta}(\bar{\mathbf{x}}) &= \left.\frac{\partial^2 U(\mathbf{x})}{\partial x_i^\alpha \partial x_j^\beta}\right|_{\mathbf{x}=\bar{\mathbf{x}}} = \frac{1}{2}\left.\frac{\partial^2 U_i(\mathbf{x}_i, \mathbf{x}_j)}{\partial x_i^\alpha \partial x_j^\beta}\right|_{\mathbf{x}=\bar{\mathbf{x}}} \\
&\quad + \frac{1}{2}\left.\frac{\partial^2 U_j(\mathbf{x}_j, \mathbf{x}_i)}{\partial x_i^\alpha \partial x_j^\beta}\right|_{\mathbf{x}=\bar{\mathbf{x}}}.
\end{aligned}
\tag{10}
$$

Similarly, the third-order tensor $\mathsf{G}(\bar{\mathbf{x}}) = \left(\mathsf{g}_{ijk}(\bar{\mathbf{x}})\right)$ $(i, j, k = 1, 2, \ldots, N)$ and the fourth-order tensor $\mathsf{Q}(\bar{\mathbf{x}}) = \left(\mathsf{q}_{ijkh}(\bar{\mathbf{x}})\right)$ $(i, j, k, h = 1, 2, \ldots, N)$ are defined respectively, where $\mathsf{g}_{ijk}(\bar{\mathbf{x}})$ and $\mathsf{q}_{ijkh}(\bar{\mathbf{x}})$ are $3 \times 3 \times 3$ and $3 \times 3 \times 3 \times 3$ tensor and their components $g_{ijk}^{\alpha\beta\gamma}(\bar{\mathbf{x}})$ $(\alpha, \beta, \gamma = 1, 2, 3)$ and $q_{ijkh}^{\alpha\beta\gamma\tau}(\bar{\mathbf{x}})$ $(\alpha, \beta, \gamma, \tau = 1, 2, 3)$ are defined by

$$
g_{ijk}^{\alpha\beta\gamma}(\bar{\mathbf{x}}) = \left.\frac{\partial^3 U(\mathbf{x})}{\partial x_i^\alpha \partial x_j^\beta \partial x_k^\gamma}\right|_{\mathbf{x}=\bar{\mathbf{x}}} = \left.\frac{\partial \mathbf{B}(\mathbf{x})}{\partial x_k^\gamma}\right|_{\mathbf{x}=\bar{\mathbf{x}}},
\tag{11}
$$

$$
q_{ijkh}^{\alpha\beta\gamma\tau}(\bar{\mathbf{x}}) = \left.\frac{\partial^4 U(\mathbf{x})}{\partial x_i^\alpha \partial x_j^\beta \partial x_k^\gamma \partial x_h^\tau}\right|_{\mathbf{x}=\bar{\mathbf{x}}} = \left.\frac{\partial^2 \mathbf{B}(\mathbf{x})}{\partial x_k^\gamma \partial x_h^\tau}\right|_{\mathbf{x}=\bar{\mathbf{x}}}.
\tag{12}
$$

### 2.3 The dynamic thermal vibration equations

The thermal vibration equations with different structural deformations and temperatures, i.e. dynamic thermal vibration equations, are established in each transient equilibrium state. The RVE considered in this work is in equilibrium state under different structural deformations and temperatures. Therefore, the velocity of structural deformation is zero, namely $\dot{\bar{\mathbf{x}}} = 0$. The kinetic energy in (3) could be expressed as

$$
K(\dot{\bar{\mathbf{x}}}) = \frac{1}{2}\dot{\bar{\mathbf{x}}}' \mathbf{M} \dot{\bar{\mathbf{x}}}.
\tag{13}
$$

In addition, each atom of RVE is in equilibrium of force. Thus, the first-order derivative of potential energy $U(\mathbf{x})$ in (7) is zero, namely

$$
\left.\frac{\partial U(\mathbf{x})}{\partial x_i^\alpha}\right|_{\mathbf{x}=\bar{\mathbf{x}}} = 0.
\tag{14}
$$

The Lagrange function is $L = K(\dot{\bar{\mathbf{x}}}) - U(\bar{\mathbf{x}} + \tilde{\mathbf{x}})$, which satisfies the following Lagrange motion equation [24,25]

$$
\frac{\mathrm{d}}{\mathrm{d}t}\left(\frac{\partial L}{\partial \dot{\tilde{\mathbf{x}}}}\right) - \frac{\partial L}{\partial \tilde{\mathbf{x}}} = 0.
\tag{15}
$$

Substituting (5) and (13) into (15) as well as applying (8)–(12) and (14), the thermal vibration equation could be obtained as follows

$$
\mathbf{M}\ddot{\tilde{\mathbf{x}}} = -\frac{\partial U(\bar{\mathbf{x}} + \tilde{\mathbf{x}})}{\partial \tilde{\mathbf{x}}} = -\mathbf{B}(\bar{\mathbf{x}})\cdot\tilde{\mathbf{x}} - \frac{1}{2}\mathsf{G}(\bar{\mathbf{x}})\cdot\tilde{\mathbf{x}}\cdot\tilde{\mathbf{x}} - \frac{1}{6}\mathsf{Q}(\bar{\mathbf{x}})\cdot\tilde{\mathbf{x}}\cdot\tilde{\mathbf{x}}\cdot\tilde{\mathbf{x}},
\tag{16}
$$

where $\ddot{\tilde{\mathbf{x}}}$ is the second-order derivative of $\tilde{\mathbf{x}}$ with respect to time $t$. $\mathbf{B}(\bar{\mathbf{x}})$ is a real symmetric matrix expressed in (8) and it has $3N$ non-negative eigenvalues $\omega_i^2(\bar{\mathbf{x}})$ $(i = 1, 2, \ldots, 3N)$, which could be obtained by the orthogonal transformation matrix $\mathbf{\Phi}(\bar{\mathbf{x}})$ with the relation $\mathbf{\Lambda}(\bar{\mathbf{x}}) = \mathbf{\Phi}'(\bar{\mathbf{x}})\mathbf{B}(\bar{\mathbf{x}})\mathbf{\Phi}(\bar{\mathbf{x}})$. The components $g_{ijk}^{\alpha\beta\gamma}(\bar{\mathbf{x}})$ of the tensor $\mathsf{G}(\bar{\mathbf{x}})$ and the $q_{ijkh}^{\alpha\beta\gamma\tau}(\bar{\mathbf{x}})$ of the tensor $\mathsf{Q}(\bar{\mathbf{x}})$ respectively expressed by (11) and (12) are described in details in "Appendix 1". By the analysis of the expressions of $g_{ijk}^{\alpha\beta\gamma}(\bar{\mathbf{x}})$ and $q_{ijkh}^{\alpha\beta\gamma\tau}(\bar{\mathbf{x}})$, it is worth noted that the diagonal terms whose subscripts satisfy $i = j = k = h$ are of great importance since they assemble the values of the other terms. Hence, we extract the diagonal terms $g_{iii}^{\alpha\beta\gamma}(\bar{\mathbf{x}})$ and $q_{iiii}^{\alpha\beta\gamma\tau}(\bar{\mathbf{x}})$ of the tensor $\mathsf{G}(\bar{\mathbf{x}})$ and $\mathsf{Q}(\bar{\mathbf{x}})$ for further analysis. While the diagonal terms $g_{iii}^{\alpha\beta\gamma}(\bar{\mathbf{x}})$ of $\mathsf{G}(\bar{\mathbf{x}})$ are zeros in the case that the RVE is in equilibrium state. More details about the proof of $g_{iii}^{\alpha\beta\gamma}(\bar{\mathbf{x}}) = 0$ are given in "Appendix 2". Further, in order to simplify calculation, we select the major terms $q_{iiii}^{\alpha\beta\gamma\tau}(\bar{\mathbf{x}})$ (i.e. $Q_{rrrr}, r = 1, 2, \cdots, 3N$) whose superscripts satisfies $\alpha = \beta = \gamma = \tau$ to study the effects of the higher-order expansion terms on thermal vibration.

The thermal vibration displacements $\tilde{\mathbf{x}}$ can be transformed to normal vibrational mode by the following normal transformation of coordinates

$$
\tilde{\mathbf{x}} = \mathbf{\Phi}(\bar{\mathbf{x}})\mathbf{s}.
\tag{17}
$$

Substituting (17) into (16) and then premultiplying $\mathbf{\Phi}'(\bar{\mathbf{x}})$, the thermal vibration equation of the $i$th vibrator with normal coordinates $\mathbf{s}$ is expressed by

$$
m_i \ddot{s}_i + \omega_i^2(\bar{\mathbf{x}}) s_i + \frac{1}{6}\lambda_i(\bar{\mathbf{x}})s_i^3 = 0, \quad i = 1, 2, \ldots, 3N,
\tag{18}
$$

where $\lambda_i(\bar{\mathbf{x}}) = \sum_{r=1}^{3N} Q_{rrrr}(\bar{\mathbf{x}})\Phi_{ri}^4(\bar{\mathbf{x}})$ is the coefficient influenced by the high-order items. The above thermal vibration equations are established at each equilibrium state of structural deformation influenced by stress and temperature. The vibration equation for each oscillator could be regard as an undamped and non-forced Duffing equation [26]. Jacobian elliptic-function method is adopted here to find its nonsingular periodic-wave solution [26–28]. Hence, the vibration function of the $i$th vibrator is

$$s_i(t) = a_i \, \text{cn} \left( \hat{\omega}_i(\bar{\mathbf{x}})t, k_i(\bar{\mathbf{x}}) \right), \tag{19}$$

where $\text{cn}\left(\hat{\omega}_i(\bar{\mathbf{x}})t, k_i(\bar{\mathbf{x}})\right)$ is Jacobian elliptic cosine function, the parameters $\hat{\omega}_i(\bar{\mathbf{x}})$ and $k_i(\bar{\mathbf{x}})$ are given by

$$\hat{\omega}_i(\bar{\mathbf{x}}) = \sqrt{\frac{\omega_i^2(\bar{\mathbf{x}})}{m_i} + \frac{\lambda_i(\bar{\mathbf{x}})a_i^2}{6m_i}},$$
$$k_i(\bar{\mathbf{x}}) = \sqrt{\frac{\lambda_i(\bar{\mathbf{x}})a_i^2}{12m_i\hat{\omega}_i^2(\bar{\mathbf{x}})}}. \tag{20}$$

The coefficient $a_i$ is vibration amplitude, which is related to the following initial conditions:

$$s_i(t_0) = 0, \quad \dot{s}_i(t_0) = \dot{\bar{x}}_i^0. \tag{21}$$

Since the total energy of the ith oscillator

$$E_i = \frac{1}{2}m_i\dot{s}_i^2 + \frac{1}{2}\omega_i^2(\bar{\mathbf{x}})s_i^2 + \frac{1}{24}\lambda_i(\bar{\mathbf{x}})s_i^4, \tag{22}$$

then by (21) and energy conservation law, we have $m_i(\dot{\bar{x}}_i^0)^2 = \omega_i^2(\bar{\mathbf{x}})a_i^2 + \lambda_i(\bar{\mathbf{x}})a_i^4/12$. Thus, the coefficient $a_i$ is expressed as

$$a_i = \sqrt{\frac{6}{\lambda_i}} \left( \sqrt{\omega_i^4(\bar{\mathbf{x}}) + \lambda_i(\bar{\mathbf{x}})m_i(\dot{\bar{x}}_i^0)^2/3} - \omega_i^2(\bar{\mathbf{x}}) \right)^{1/2}. \tag{23}$$

The period of vibration function (19) is $\Gamma_i = \frac{4}{\hat{\omega}_i(\bar{\mathbf{x}})} \int_0^{\pi/2} \left(1 - k_i^2(\bar{\mathbf{x}})\sin^2\varphi\right)^{-\frac{1}{2}} d\varphi$. By Taylor expansion, the period $\Gamma_i$ can be written as

$$\Gamma_i = \frac{4}{\hat{\omega}_i(\bar{\mathbf{x}})} \int_0^{\pi/2} \left(1 + \frac{1}{2}k_i^2(\bar{\mathbf{x}})\sin^2\varphi + O(k_i^4(\bar{\mathbf{x}})\sin^4\varphi)\right) d\varphi. \tag{24}$$

The thermal vibration is of high frequency and small amplitude. And the amplitude is a small quantity relative to the structural deformation position and it satisfy $a_i \ll 1$. Therefore, the relation $k_i^2(\bar{\mathbf{x}}) \ll 1$ is easy to obtained according to (20). Therefore, we drop the terms that higher than $k_i^2(\bar{\mathbf{x}})\sin^2\varphi$ in (24) in the next calculation. Then the period above can be expressed approximately by $\Gamma_i \approx 2\pi(1 + \frac{1}{4}k_i^2(\bar{\mathbf{x}}))/\hat{\omega}_i(\bar{\mathbf{x}})$. Besides, $\hat{\omega}_i(\bar{\mathbf{x}}) \approx \omega_i(\bar{\mathbf{x}})(1 + k_i^2(\bar{\mathbf{x}}))/\sqrt{m_i}$. Therefore, the dynamic thermal vibration frequency of the $i$th vibrator is

$$\varpi_i^d(\bar{\mathbf{x}}, T) = \frac{2\pi}{\Gamma_i}$$
$$= \frac{\omega_i(\bar{\mathbf{x}})}{\sqrt{m_i}} \left(1 + \frac{\lambda_i(\bar{\mathbf{x}})a_i^2}{16\left(\omega_i^2(\bar{\mathbf{x}}) + \lambda_i(\bar{\mathbf{x}})a_i^2/6\right)}\right),$$
$$i = 1, 2, \dots, 3N. \tag{25}$$

The superscript $d$ denotes that the thermal vibration frequency of each vibrator dynamically depend on the structural deformation $\bar{\mathbf{x}}(f, T)$ and the temperature $T$. Based on dynamic thermal vibration, the heat capacity and thermal expansion coefficient will be obtained in the next section.

# 3 Heat capacity and TEC of metal crystalline materials

In this section, according to the statistical thermodynamics [3,7,24], the heat capacity and thermal expansion coefficient (TEC) for metal crystalline materials are derived from the dynamic thermal vibration frequencies.

## 3.1 The formula of heat capacity

The energy $E_{n_s}$ in $n_s$ level is given by

$$E_{n_s}(\bar{\mathbf{x}}, T) = U_0(\bar{\mathbf{x}}) + \sum_{i=1}^{3N} \left( \left(n_s + \frac{1}{2}\right)\hbar\varpi_i^d \right),$$
$$n_s = 0, 1, 2, \dots, \tag{26}$$

where $U_0(\bar{\mathbf{x}})$ is the energy expressed by (6), $\varpi_i^d = \varpi_i^d(\bar{\mathbf{x}}, T)$ is the dynamic thermal vibration frequency of the $i$th vibrator shown in (25) and $\hbar$ is Planck constant. Then the partition function of RVE is shown as follows

$$Z = \sum_{n_s} e^{-\beta E_{n_s}(\bar{\mathbf{x}}, T)} = e^{-\beta U_0(\bar{\mathbf{x}})} \prod_{i=1}^{3N} \frac{e^{-\hbar\varpi_i^d/2k_BT}}{1 - e^{-\hbar\varpi_i^d/k_BT}}, \tag{27}$$

where $\beta = 1/k_BT$ and $k_B$ is Boltzmann constant.

Before deriving the formula of heat capacity for metal crystalline materials from the partition function, we introduce the following temperature-dependent parameter $\delta_i$ to describe how the thermal vibration frequency with temperature

$$\delta_i = \delta_i(\bar{\mathbf{x}}, T) = -\frac{\partial \ln \varpi_i^d(\bar{\mathbf{x}}, T)}{\partial \ln T}, i = 1, 2, \dots, 3N. \tag{28}$$

The first-order derivative of $\delta_i$ with respect to temperature is

$$\frac{\partial \delta_i}{\partial T} = -\frac{T}{\varpi_i^d} \frac{\partial^2 \varpi_i^d}{\partial T^2} + \frac{\delta_i}{T} + \frac{\delta_i^2}{T}. \tag{29}$$

For metal crystalline materials, the thermal vibration frequency have an approximate linear relationship with the temperature in the range of $0\,\mathrm{K}$ to the melting point, which has been validated by Fig. 2 in numerical results of Sect. 5. Thus, we obtain $\partial^2 \varpi_i^d / \partial T^2 \approx 0$, it follows that

$$\frac{\partial \delta_i}{\partial T} \approx \frac{\delta_i}{T} + \frac{\delta_i^2}{T}. \tag{30}$$

Besides, according to (28) and the expression $T = 1/k_B \beta$, we have

$$\frac{\partial}{\partial \beta} \left( \frac{\hbar \varpi_i^d}{k_B T} \right) = (1 + \delta_i) \hbar \varpi_i^d. \tag{31}$$

Then the internal energy $\bar{E}(\bar{\mathbf{x}}, T)$ can be obtained by (27), (28), (31) as follows

$$\begin{aligned}
\bar{E}(\bar{\mathbf{x}}, T) &= -\frac{\partial}{\partial \beta} \ln Z \\
&= U_0(\bar{\mathbf{x}}) + \sum_{i=1}^{3N} (1 + \delta_i) \left( \frac{1}{2} \hbar \varpi_i^d + \frac{\hbar \varpi_i^d}{e^{\hbar \varpi_i^d / k_B T} - 1} \right) \\
&\approx U_0(\bar{\mathbf{x}}) + (1 + \bar{\delta}) \sum_{i=1}^{3N} \left( \frac{1}{2} \hbar \varpi_i^d + \frac{\hbar \varpi_i^d}{e^{\hbar \varpi_i^d / k_B T} - 1} \right),
\end{aligned} \tag{32}$$

where $\bar{\delta}$ is the average value of $\delta_i$ ($i = 1, 2, \ldots, 3N$) and is expressed as

$$\bar{\delta} = \bar{\delta}(\bar{\mathbf{x}}, T) = \frac{1}{3N} \sum_{i=1}^{3N} \delta_i(\bar{\mathbf{x}}, T). \tag{33}$$

Further, the frequency distribution function $f(\omega)$ in Debye model [1–3,7] is adopted here,

$$f(\omega) = \begin{cases} l\omega^2, & \omega \in [\omega_{\min}, \omega_{\max}], \\ 0, & \omega \in [0, \omega_{\min}) \cup (\omega_{\max}, +\infty), \end{cases} \tag{34}$$

where $l = 9N/(\omega_{\max}^3 - \omega_{\min}^3)$ is the normalization constant; $\omega_{\min}$ and $\omega_{\max}$ are

$$\omega_{\min} = \min_{1 \le i \le 3N} \{\varpi_i^d(\bar{\mathbf{x}}, T)\}, \quad \omega_{\max} = \max_{1 \le i \le 3N} \{\varpi_i^d(\bar{\mathbf{x}}, T)\}. \tag{35}$$

Besides, we denote

$$\theta_i^d = \theta_i^d(\bar{\mathbf{x}}, T) = \frac{\hbar \varpi_i^d}{k_B} \in [\theta_{\min}, \theta_{\max}], \tag{36}$$

where $\theta_{\min} = \hbar \omega_{\min}/k_B$ and $\theta_{\max} = \hbar \omega_{\max}/k_B$. By applying (34)–(36), the internal energy (32) can be approximately rewritten as the following integral form

$$\begin{aligned}
\bar{E}(\bar{\mathbf{x}}, T) &= U_0(\bar{\mathbf{x}}) \\
&\quad + (1 + \bar{\delta}) \int_{\omega_{\min}}^{\omega_{\max}} \left( \frac{1}{2} \hbar \omega + \frac{\hbar \omega}{e^{\hbar \omega / k_B T} - 1} \right) f(\omega) \, d\omega \\
&= U_0(\bar{\mathbf{x}}) + (1 + \bar{\delta}) c k_B T^4 \int_{\theta_{\min}/T}^{\theta_{\max}/T} \left( \frac{1}{2} + \frac{1}{e^\zeta - 1} \right) \zeta^3 \, d\zeta,
\end{aligned} \tag{37}$$

where $c = 9N/(\theta_{\min}^3 - \theta_{\max}^3)$ and $\zeta = \hbar \omega / k_B T$. Furthermore, the maximum value $\theta_{\max}$ is recognized as Debye temperature [2,7], namely $\Theta_D = \theta_{\max}$.

According to the internal energy $\bar{E}(\bar{\mathbf{x}}, T)$ in (32), the frequency distribution function in (34), the expressions (28), (30) and (36), we can obtain the heat capacity $C_V$ at constant volume and the heat capacity $C_P$ at constant pressure as follows

$$C_V = \left( \frac{\partial \bar{E}(\bar{\mathbf{x}}, T)}{\partial T} \right)_V = c k_B T^3 \int_{\theta_{\min}/T}^{\theta_{\max}/T} \frac{e^\zeta}{(e^\zeta - 1)^2} \zeta^4 \, d\zeta, \tag{38}$$

$$C_P = \left( \frac{\partial (\bar{E}(\bar{\mathbf{x}}, T) + PV)}{\partial T} \right)_P = (1 + \bar{\delta})^2 C_V + PV\alpha_V, \tag{39}$$

where $P$ is the pressure, $V$ is the volume and $\alpha_V$ is the TEC that defined by (45).

## 3.2 The formula of TEC

By the partition function in (27) and the expression (36), the Helmholtz free energy is expressed as

$$\begin{aligned}
F(\bar{\mathbf{x}}, T) &= -k_B T \ln Z \\
&= U_0(\bar{\mathbf{x}}) + k_B T \sum_{i=1}^{3N} \left( \frac{1}{2} \frac{\theta_i^d}{T} + \ln \left( 1 - e^{-\theta_i^d / T} \right) \right).
\end{aligned} \tag{40}$$

According to the dynamic thermal vibration frequency $\varpi_i^d(\bar{\mathbf{x}}, T)$ in (25) and Ref. [4,5], the Grüneisen parameter $\gamma_i(\bar{\mathbf{x}}, T)$ is defined by

$$\gamma_i(\bar{\mathbf{x}}, T) = -\frac{\partial \ln \varpi_i^d(\bar{\mathbf{x}}, T)}{\partial \ln V}. \tag{41}$$

It should be pointed out that Grüneisen parameter depend on both the temperature $T$ and structural deformation position $\bar{\mathbf{x}}$ in this work. Besides, the first-order derivative of $\theta_i^d$ with respect to the volume is $(\partial \theta_i^d / \partial V)_T = \gamma_i(\bar{\mathbf{x}}, T)\theta_i^d / V$. Then the pressure can be obtained with (40) and (41) as follows

$$
\begin{aligned}
P(V, T) = & -\left(\frac{\partial F(\bar{\mathbf{x}}, T)}{\partial V}\right)_T = -\left(\frac{dU_0(\bar{\mathbf{x}})}{dV}\right)_T \\
& + k_B \frac{\bar{\gamma}}{V} \sum_{i=1}^{3N} \theta_i^d \left(\frac{1}{2} + \frac{1}{e^{\theta_i^d / T} - 1}\right),
\end{aligned} \tag{42}
$$

where $\bar{\gamma}$ is the average value of $\gamma_i$ ($i = 1, 2, \ldots, 3N$) and is expressed as

$$
\bar{\gamma} = \bar{\gamma}(\bar{\mathbf{x}}, T) = \frac{1}{3N} \sum_{i=1}^{3N} \gamma_i(\bar{\mathbf{x}}, T). \tag{43}
$$

When the pressure $P(V, T)$ remains a constant, there exists the relation

$$
\left(\frac{\partial P}{\partial T}\right)_V + \left(\frac{\partial P}{\partial V}\right)_T \cdot \left(\frac{\partial V}{\partial T}\right)_P = 0. \tag{44}
$$

Thus, it is easy to obtain the formula of TEC as follows

$$
\alpha_V = \frac{1}{V}\left(\frac{\partial V}{\partial T}\right)_P = -\frac{1}{V}\frac{(\partial P / \partial T)_V}{(\partial P / \partial V)_T}. \tag{45}
$$

According to the numerical results shown in Fig. 3, the thermal vibration frequency change linearly with the volume approximately, namely $\partial^2 \varpi_i^d / \partial V^2 \approx 0$. Thus,

$$
\frac{\partial \bar{\gamma}}{\partial V} = -\frac{V}{\varpi_i^d} \frac{\partial^2 \varpi_i^d}{\partial V^2} + \frac{\bar{\gamma}}{V} + \frac{\bar{\gamma}^2}{V} \approx \frac{\bar{\gamma}}{V} + \frac{\bar{\gamma}^2}{V} \tag{46}
$$

Then substituting (42) into (45) and applying (46), we have

$$
\left(\frac{\partial P}{\partial T}\right)_V = \frac{\bar{\gamma}C_V}{V}, \quad \left(\frac{\partial P}{\partial V}\right)_T = \frac{\bar{\gamma}^2 T C_V}{V^2} - \left(\frac{\partial^2 U_0(\bar{\mathbf{x}})}{\partial V^2}\right)_T, \tag{47}
$$

where $C_V$ is the heat capacity defined by (38). Therefore, by substituting (47) into (45), the formula of volumetric TEC can be written as

$$
\alpha_V = \frac{\bar{\gamma}C_V}{B_0(\bar{\mathbf{x}})V - \bar{\gamma}^2 T C_V}, \tag{48}
$$

where $B_0(\bar{\mathbf{x}}) = V \cdot (\partial^2 U_0(\bar{\mathbf{x}}) / \partial V^2)_T$ is the bulk modulus. The linear thermal expansion coefficient $\alpha_l$ satisfies $\alpha_l = \alpha_V / 3$ for isotropic material.

# 4 Algorithm procedure

The detailed algorithm procedure for predicting the heat capacity and TEC of metal crystalline materials is stated as follows:

(i) Construct an initial lattice configuration with the lattice constant $a_0$ (shown in Table 1) as the RVE; The ERVE is chosen as 1.5–3.0 times the RVE.
(ii) Relax the ERVE by LAMMPS software with the NPT ensemble at $T = T_0$ with the pressure $P = 0$ and compute the current temperature $T_{cur}$ until $T_{cur}$ is steadily near the temperature $T$;
(iii) Export the data of current configuration at $T$ and the velocities at $T$ from LAMMPS; Then record the structure deformation positions $\bar{\mathbf{x}}(T)$ and the velocity $\dot{\bar{\mathbf{x}}}^0$ of all atoms in RVE and ERVE;
(iv) Build the matrix $\mathbf{B}(\bar{\mathbf{x}})$ according to (8)–(10) and the tensor $\mathbf{Q}(\bar{\mathbf{x}})$ expressed by (12) at the structural deformation position $\bar{\mathbf{x}}(T)$, respectively.
(v) Compute $3N$ eigenvalues $\omega_i^2(\bar{\mathbf{x}})$, $i = 1, 2, \ldots, 3N$ at $T$ by the standard orthogonal transformation matrix $\mathbf{\Phi}(\bar{\mathbf{x}})$ and then the $3N$ dynamic thermal vibration frequency $\varpi_i^d(\bar{\mathbf{x}}(T), T)$ expressed by (25) are obtained;
(vi) Set $T = T_0 + \triangle T$ and go to the Step (ii) to recalculate $\varpi_i^d(\bar{\mathbf{x}}(T), T)$ until the temperature $T$ is slightly higher than the melting point, where $\triangle T$ is the temperature step;
(vii) Calculate the parameter $\delta_i(\bar{\mathbf{x}}, T)$ and $\gamma_i(\bar{\mathbf{x}}, T)$ of $3N$ vibrators by (28) and (41) at different temperatures. Therefore, the heat capacity $C_V$ and $C_P$, and the TEC $\alpha_V$ at different temperatures could be obtained by (38), (39), (45), respectively.

# 5 Numerical examples

The heat capacity and TEC for face-centered cubic (fcc) metals Copper (Cu), Aluminum (Al), Gold (Au), Silver (Ag), Nickel (Ni), Palladium (Pd), Platinum(Pt) and Plumbum (Pb) from 0 K to the melting point are calculated to validate our approach. The material parameters of these metals are shown in Table 1. The RVE is set as $14.46\,\text{Å} \times 14.46\,\text{Å} \times 14.46\,\text{Å}$, which contains 64 lattices. And the size of ERVE is $18.075\,\text{Å} \times 18.075\,\text{Å} \times 18.075\,\text{Å}$. The temperature changes from $T_0 = 0$ to $T_{end}$ for each metal with $\triangle T = 10\,\text{K}$, where the temperature $T_{end}$ is a multiply of $\triangle T$ and is slightly higher than the melting point. Besides, the potential energy for the metal materials is chosen as EAM potential energy and it is expressed as follows [29–32]

**Table 1** The material parameters of eight metals

| Metals | Cu | Al | Au | Ag | Ni | Pd | Pt | Pb |
|---|---|---|---|---|---|---|---|---|
| $a_0(\text{Å})$ | 3.615 | 4.045 | 4.073 | 4.073 | 3.521 | 3.889 | 3.917 | 4.950 |
| $M(u)$ | 63.55 | 26.98 | 196.97 | 107.87 | 58.71 | 106.40 | 195.09 | 207.19 |
| $M.P.(K)$ | 1357.6 | 933.5 | 1337.6 | 1235.1 | 1728.0 | 1827.0 | 2045.0 | 600.6 |

$a_0$ is the lattice constant, $M.P.$ is the melting point, $M$ denotes the relative atomic mass, $1u = 1.66083 \times 10^{-27}$ kg

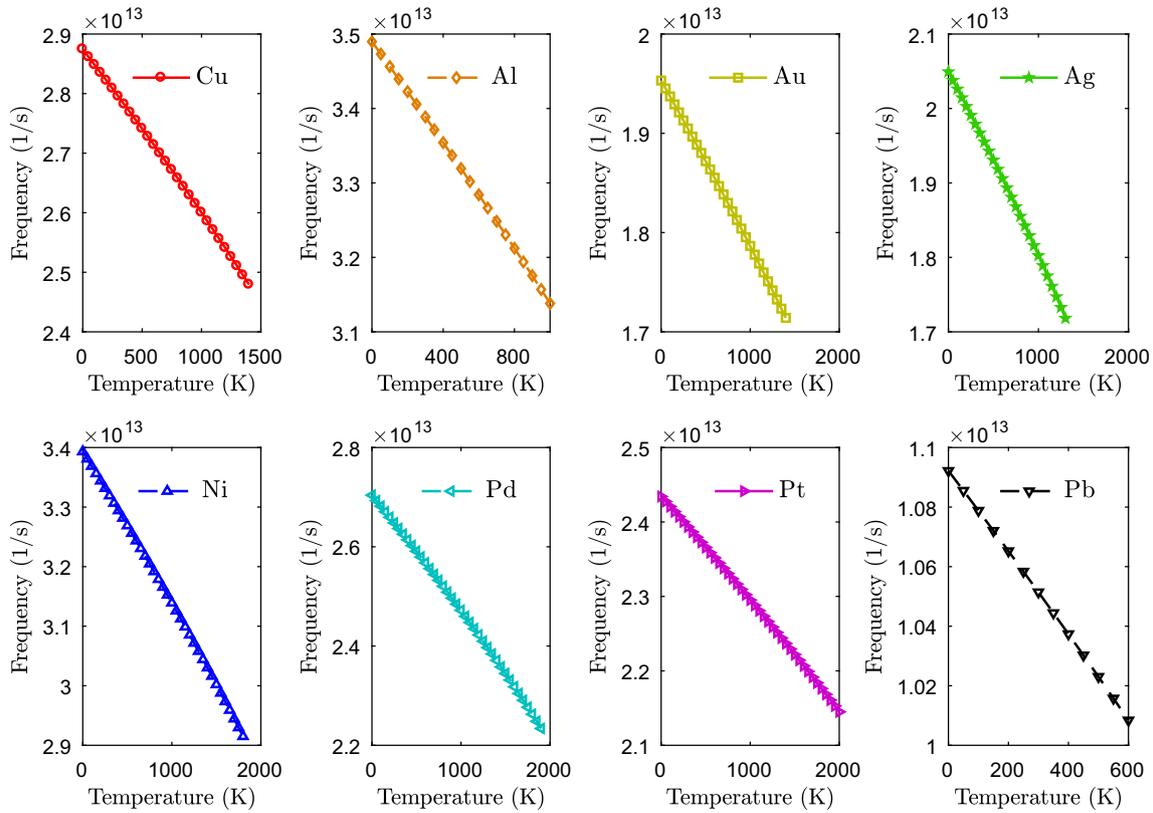

**Fig. 2** The variation tendency of thermal vibration frequency with respect to the temperature from 0 K to the melting point for single atom of Cu, Al, Au, Ag, Ni, Pd, Pt and Pb (Color online)

$$U = \sum_i \left( F\left( \sum_{j \neq i} \rho(r_{ij}) \right) + \frac{1}{2} \sum_{j \neq i} V(r_{ij}) \right), \quad (49)$$

where $r_{ij} = |\mathbf{x}_i - \mathbf{x}_j|$ is the distance between the $i$th and $j$th atom, $\bar{\rho}_i = \sum_{j \neq i} \rho(r_{ij})$ is the electron cloud density, $V(r_{ij})$ and $F(\bar{\rho}_i)$ are the pair potential energy and embedded atom energy, respectively.

Figures 2 and 3 show the thermal vibration frequency of Cu, Al, Au, Ag, Ni, Pd, Pt and Pb at different temperatures and volumes. It is obvious that the frequency have approximate linear relationship with the temperature as well as volume.

Figures 4 and 5 display the values of the parameter $\bar{\delta}(\bar{\mathbf{x}}, T)$ and Grüneisen parameter $\bar{\gamma}(\bar{\mathbf{x}}, T)$ of Cu, Al, Au, Ag, Ni, Pd, Pt and Pb from 0 K to the melting point, respectively. It is found that both the parameter $\bar{\delta}(\bar{\mathbf{x}}, T)$ and $\bar{\gamma}(\bar{\mathbf{x}}, T)$ have an approximate linear relationship with temperature. Therefore, the polynomial fitting formulae for $\bar{\delta}(\bar{\mathbf{x}}, T)$ and $\bar{\gamma}(\bar{\mathbf{x}}, T)$ are shown as follows

$$\begin{aligned} \bar{\delta}(T) &= p_0 T, \\ \bar{\gamma}(T) &= q_0 + q_1 T. \end{aligned} \quad (50)$$

Table 2 displays the fitting coefficients $p_0$, $q_0$ and $q_1$ of $\bar{\delta}(\bar{\mathbf{x}}, T)$ and $\bar{\gamma}(\bar{\mathbf{x}}, T)$ for metals Cu, Al, Au, Ag, Ni, Pd, Pt and Pb.

Figure 6 illustrates the numerical results of heat capacity for Cu, Al, Au, Ag, Ni, Pd, Pt and Pb from 0 K to the melting point in present work, comparing with the data under harmonic approximation [1,2], quasi-harmonic approximation [4,5] and experimental data [33]. The results of heat capacity show a temperature dependence. At temperatures lower than Debye temperature, the harmonic data, quasi-harmonic data and the data of present work could match the experimental data well. However, as temperature grows to the melting point, the harmonic data become constant and the quasi-harmonic data deviate from the experimental data gradually. While the data of present work are in good agreement with the experimental data both at temperatures lower and higher than Debye temperature, up to the melting point.

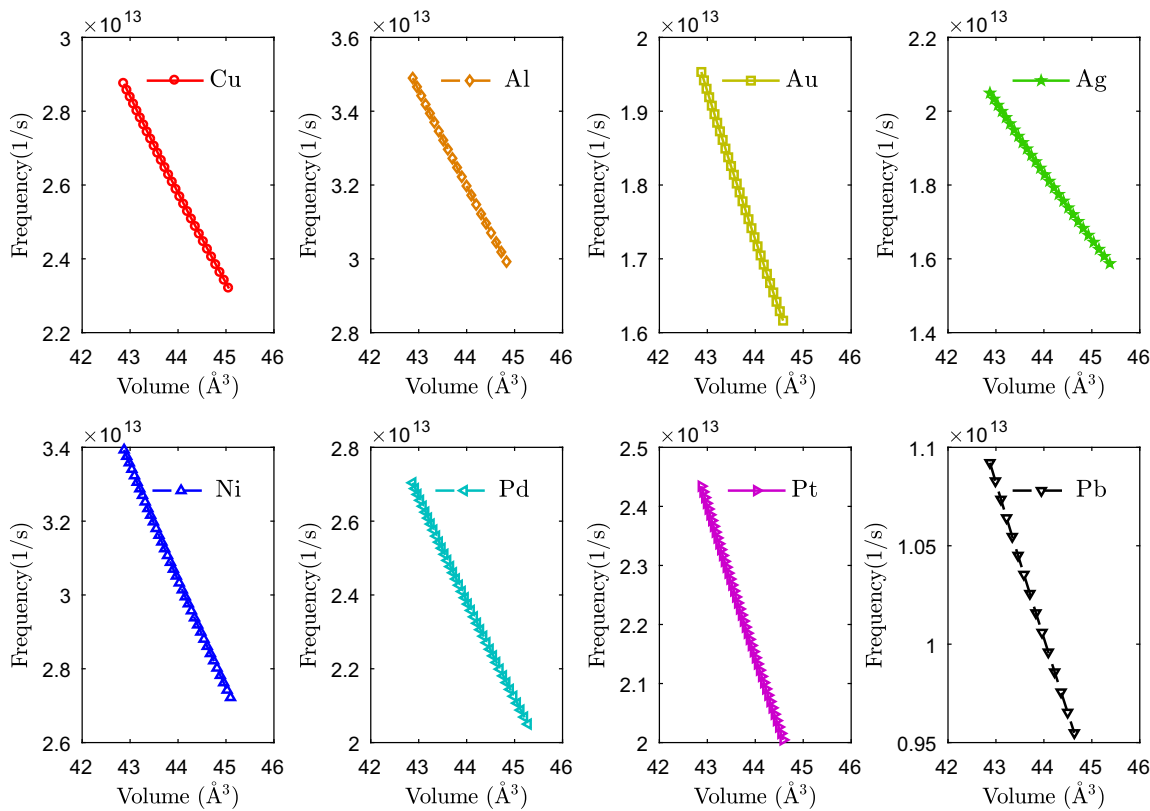

**Fig. 3** The variation tendency of thermal vibration frequency with respect to volume of one lattice for Cu, Al, Au, Ag, Ni, Pd, Pt and Pb (Color online)

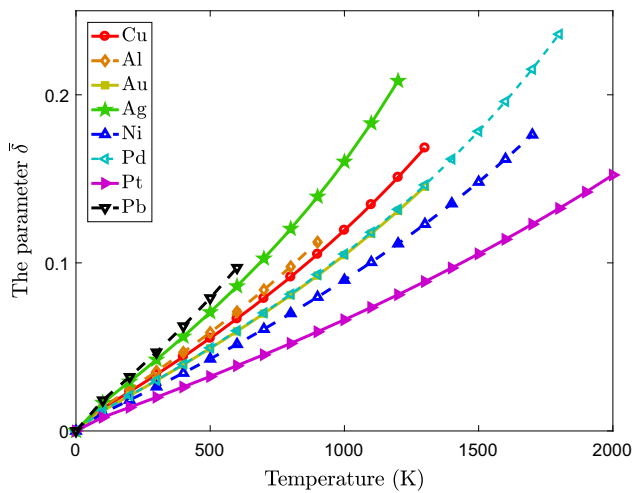

**Fig. 4** The parameter $\bar{\delta}$ of metals Cu, Al, Au, Ag, Ni, Pd, Pt and Pb from 0 K to the melting points (Color online)

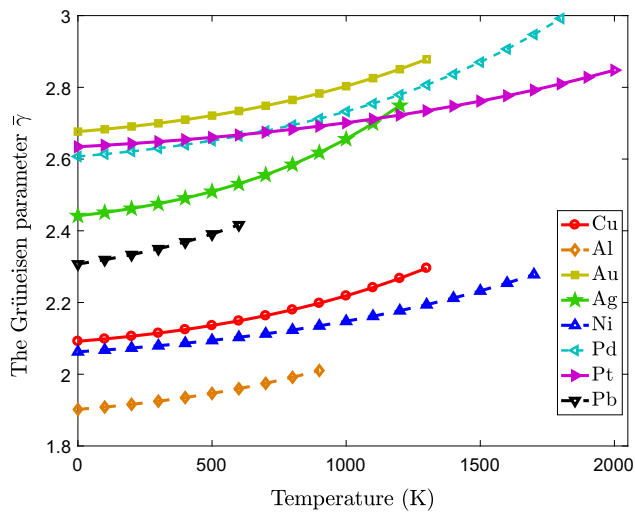

**Fig. 5** The Grüneisen parameter $\bar{\gamma}$ of metals Cu, Al, Au, Ag, Ni, Pd, Pt and Pb from 0 K to the melting points (Color online)

**Table 2** The fitting coefficients for $\bar{\gamma}$ and $\bar{\delta}$

| The fitting coefficients | Cu | Al | Au | Ag | Ni | Pd | Pt | Pb |
|---|---|---|---|---|---|---|---|---|
| $p_0(\times 10^{-4})$ | 1.12 | 1.21 | 1.05 | 1.51 | 0.95 | 1.05 | 0.71 | 1.63 |
| $q_0$ | 2.09 | 1.90 | 2.67 | 2.44 | 2.06 | 2.60 | 2.63 | 2.30 |
| $q_1 (\times 10^{-4})$ | 1.43 | 1.00 | 1.50 | 2.31 | 1.20 | 2.02 | 1.00 | 2.01 |

**Fig. 6** Comparison of the harmonic data [1,2], quasi-harmonic data [4,5], experimental data [33] and the present work for the heat capacity at constant pressure from 0 K to the melting points: **a** Cu; **b** Al; **c** Au; **d** Ag; **e** Ni; **f** Pd; **g** Pt; **h** Pb (Color online)

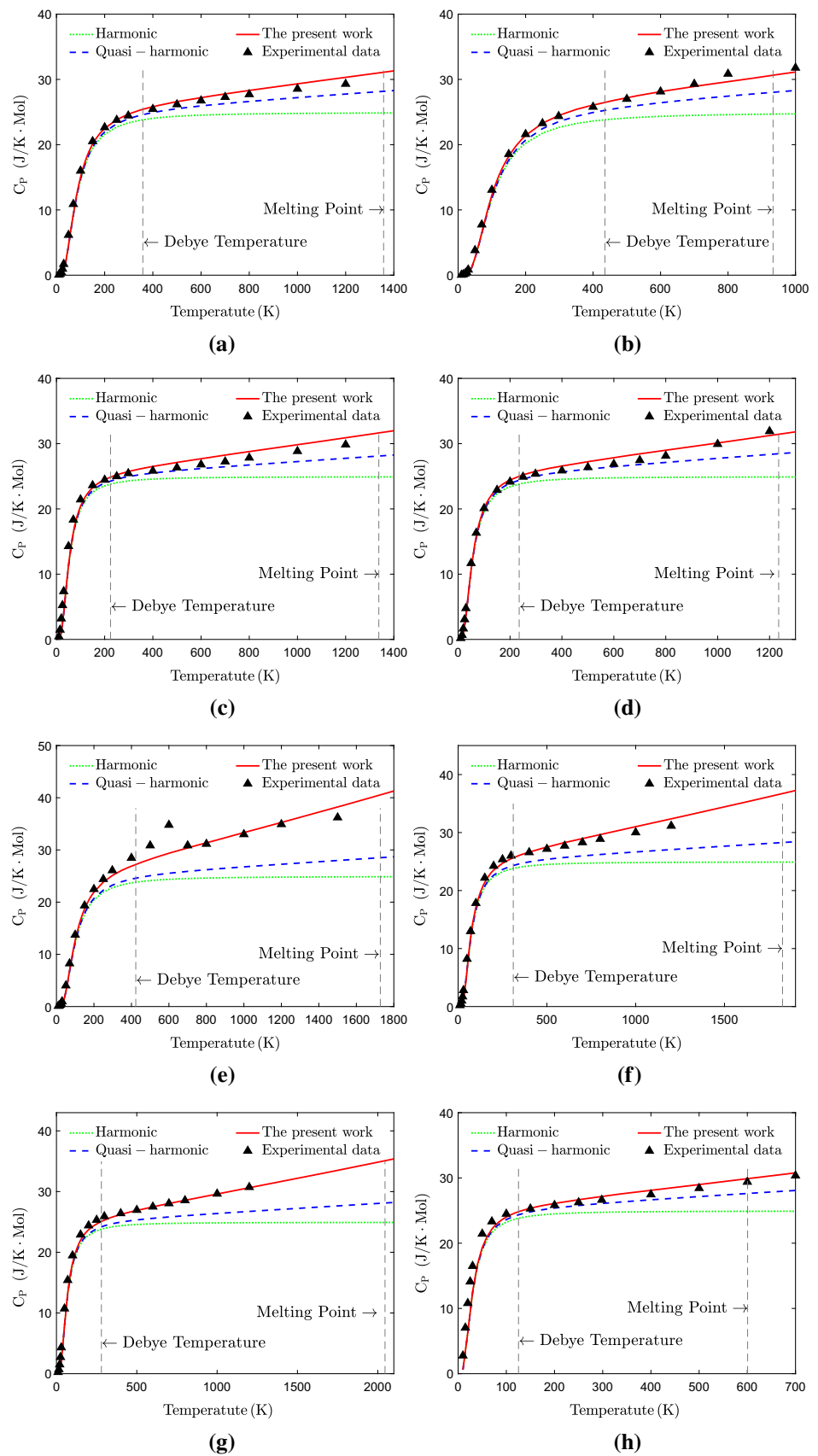

**Fig. 7** Comparison of the quasi-harmonic data [4,5], experimental data [33,34] and the present work for the TEC from 0 K to the melting points: **a** Cu; **b** Al; **c** Au; **d** Ag; **e** Ni; **f** Pd; **g** Pt; **h** Pb (Color online)

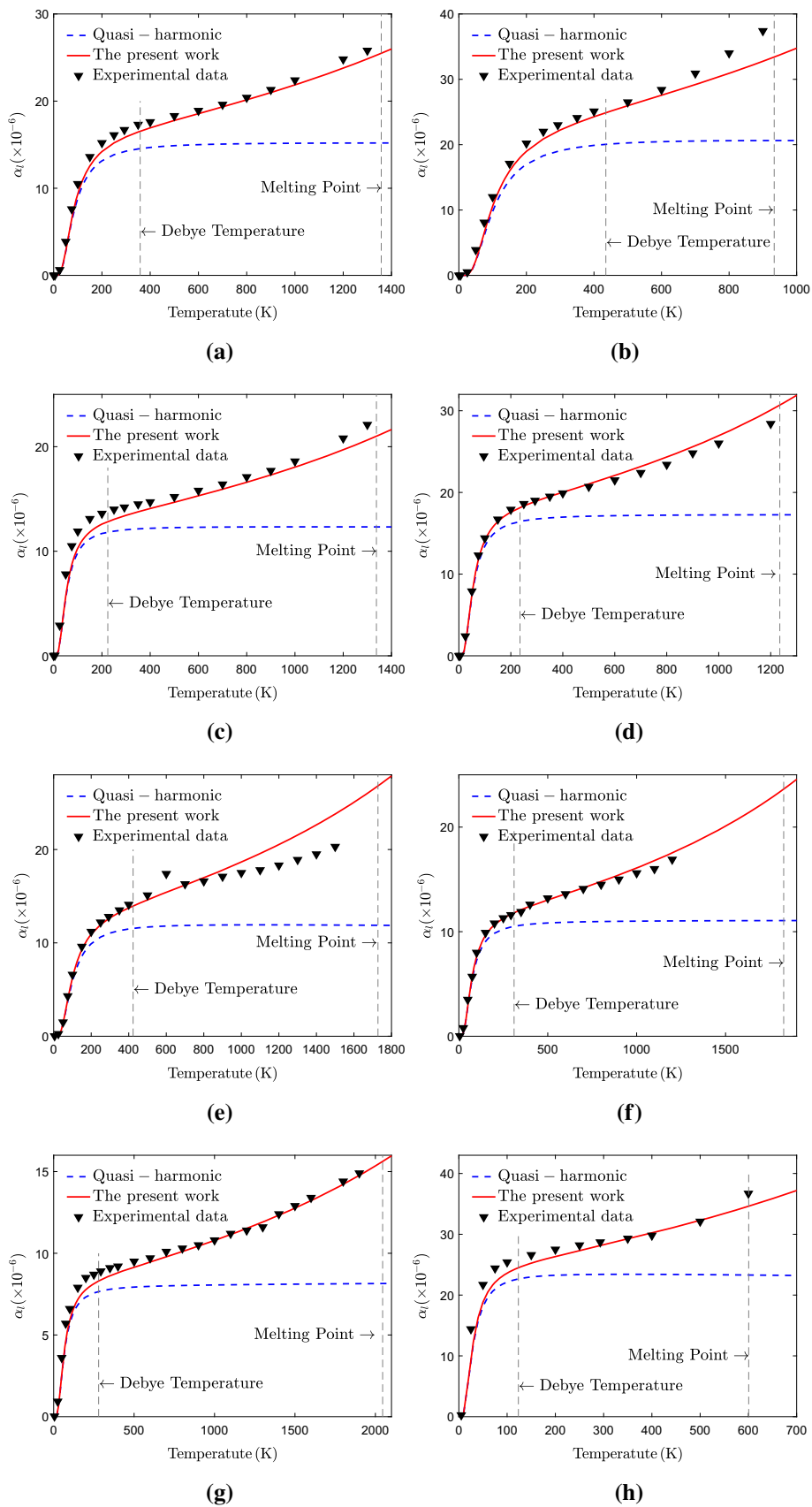

Figure 7 displays the numerical results of TEC for Cu, Al, Au, Ag, Ni, Pd, Pt and Pb from 0 K to the melting point in present work, comparing with the data under quasi-harmonic approximation [4,5] and the experimental data [33,34]. The linear TEC has an analogous tendency as heat capacity. Similarly, the superiority of our approach is validated again by TEC. The numerical results of TEC based on our approach show better performance than the quasi-harmonic data, especially at the temperature higher than Debye temperature.

As can be observed from Figs. 6 and 7, both the results of heat capacity and TEC of our approach are superior to the data under harmonic or quasi-harmonic approximation. The thermal vibration frequency obtained by (25) dynamically depend on the structural deformation and the temperature. Both the parameter $\bar{\gamma}$ and $\bar{\delta}$ are temperature-dependent. Thus, at each temperature stage, the heat capacity and TEC that derived from the dynamic thermal vibration frequencies and the parameter $\bar{\gamma}$ as well as $\bar{\delta}$ match the experimental data.

# 6 Conclusions

In this paper, we study the heat capacity and thermal expansion coefficient (TEC) of metal crystalline materials by considering the dynamic thermal vibration. The Taylor expansion terms of Hamiltonian up to fourth-order are applied to establish the thermal vibration equations. It is worth to emphasize that the expansion points are at the transient structural deformation positions which influenced by the temperature and the stress. The thermal vibration frequencies are obtained by solving thermal vibration equations through Jacobi elliptic-function method. As a conclusion, the thermal vibration frequencies dynamically change with the structural deformation and the temperature. The relation of the thermal vibration frequency with the temperature plays a vital role to illustrate heat capacity and TEC in the case of high temperatures. Further, the parameter $\bar{\delta}(\bar{x}, T)$ and Grüneisen parameter $\bar{\gamma}(\bar{x}, T)$, which show a linear correlation of the temperature, are introduced to derive the formulae of heat capacity and TEC. The numerical results for fcc metals Cu, Al, Au, Ag, Ni, Pd, Pt and Pb are in good agreement with the experimental data from 0 K to the melting point, relative to the harmonic data and quasi-harmonic data. The approach developed in this paper could be extended to investigate the thermodynamic properties of other metal crystalline materials or their alloys with different lattice structures.

**Acknowledgements** This research was financially supported by the National Key Research and Development Program of China (2016YFB1 100602), National Natural Science Foundation of China (51739007, 11501449), the Fundamental Research Funds for the Central Universities (3102017zy043), and the China Postdoctoral Science Foundation (2018M633569).

# Appendix 1: The components of the tensor $\mathsf{G}(\bar{x})$ and $\mathsf{Q}(\bar{x})$

In this appendix, we present the detailed expressions of the component $g_{ijk}^{\alpha\beta\gamma}(\bar{x})$ and $q_{ijkh}^{\alpha\beta\gamma\tau}(\bar{x})$ for tensor $\mathsf{G}(\bar{x})$ and $\mathsf{Q}(\bar{x})$, according to the definition expressed by (11) and (12).

The total potential energy $U(\mathbf{x})$ in (4) can be written as follows

$$U(\mathbf{x}) = \frac{1}{2} \sum_{i=1}^{N} \sum_{j=1, j\neq i}^{N_E} U_i(r_{ij}), \tag{A.1}$$

where $\mathbf{x} = (x_1^1, x_1^2, x_1^3, \ldots, x_N^1, x_N^2, x_N^3)$ and $r_{ij} = \left((x_i^1 - x_j^1)^2 + (x_i^2 - x_j^2)^2 + (x_i^3 - x_j^3)^2\right)^{1/2}$ is the distance between the $i$th and $j$th atoms.

By (A.1) and (11), the components $g_{ijk}^{\alpha\beta\gamma}(\bar{x})$ of tensor $\mathsf{G}(\bar{x})$ are presented by the following three cases:

(i) $i = j = k$, $i = 1, 2, \ldots, N$ and $\alpha, \beta, \gamma = 1, 2, 3$.

$$g_{iii}^{\alpha\beta\gamma}(\bar{x}) = \frac{1}{2} \sum_{p=1, p\neq i}^{N_E} \frac{\partial^3 U_i(r_{ip})}{\partial x_i^\alpha \partial x_i^\beta \partial x_i^\gamma}\bigg|_{\mathbf{x}=\bar{\mathbf{x}}} + \frac{1}{2} \sum_{l=1, l\neq i}^{N} \frac{\partial^3 U_l(r_{li})}{\partial x_i^\alpha \partial x_i^\beta \partial x_i^\gamma}\bigg|_{\mathbf{x}=\bar{\mathbf{x}}}. \tag{A.2}$$

(ii) $i = j \neq k$, $i, k = 1, 2, \ldots, N$ and $\alpha, \beta, \gamma = 1, 2, 3$.

$$g_{iik}^{\alpha\beta\gamma}(\bar{x}) = g_{iki}^{\alpha\beta\gamma}(\bar{x}) = g_{kii}^{\alpha\beta\gamma}(\bar{x})$$
$$= \frac{1}{2} \frac{\partial^3 U_i(r_{ik})}{\partial x_i^\alpha \partial x_i^\beta \partial x_k^\gamma}\bigg|_{\mathbf{x}=\bar{\mathbf{x}}} + \frac{1}{2} \frac{\partial^3 U_k(r_{ki})}{\partial x_i^\alpha \partial x_i^\beta \partial x_k^\gamma}\bigg|_{\mathbf{x}=\bar{\mathbf{x}}}. \tag{A.3}$$

(iii) $i \neq j \neq k$, $i, j, k = 1, 2, \ldots, N$ and $\alpha, \beta, \gamma = 1, 2, 3$.

$$g_{ijk}^{\alpha\beta\gamma}(\bar{x}) = 0. \tag{A.4}$$

Similarly, with (A.1) and (12), the components $q_{ijkh}^{\alpha\beta\gamma\tau}(\bar{x})$ of tensor $\mathsf{Q}(\bar{x})$ can be shown as the following three cases:

(i) $i = j = k = h$, $i = 1, 2, \ldots, N$ and $\alpha, \beta, \gamma, \tau = 1, 2, 3$.

$$
\begin{aligned}
q_{iiii}^{\alpha\beta\gamma\tau}(\bar{\mathbf{x}}) &= \frac{\partial^4 U(\mathbf{x})}{\partial x_i^\alpha \partial x_i^\beta \partial x_i^\gamma \partial x_i^\tau}\bigg|_{\mathbf{x}=\bar{\mathbf{x}}} \\
&= \frac{1}{2} \sum_{p=1, p \neq i}^{N_E} \frac{\partial^4 U_i(r_{ip})}{\partial x_i^\alpha \partial x_i^\beta \partial x_i^\gamma \partial x_i^\tau}\bigg|_{\mathbf{x}=\bar{\mathbf{x}}} \\
&\quad + \frac{1}{2} \sum_{l=1, l \neq i}^{N} \frac{\partial^4 U_l(r_{li})}{\partial x_i^\alpha \partial x_i^\beta \partial x_i^\gamma \partial x_i^\tau}\bigg|_{\mathbf{x}=\bar{\mathbf{x}}}.
\end{aligned}
\tag{A.5}
$$

(ii) $i = j = k \neq h$ or $i = j \neq k = h$, $i, h = 1, 2, \ldots, N$ and $\alpha, \beta, \gamma, \tau = 1, 2, 3$.

$$
\begin{aligned}
q_{iiih}^{\alpha\beta\gamma\tau}(\bar{\mathbf{x}}) &= q_{iihi}^{\alpha\beta\gamma\tau}(\bar{\mathbf{x}}) = q_{ihii}^{\alpha\beta\gamma\tau}(\bar{\mathbf{x}}) = q_{hiii}^{\alpha\beta\gamma\tau}(\bar{\mathbf{x}}) \\
&= \frac{1}{2} \frac{\partial^4 U_i(r_{ih})}{\partial x_i^\alpha \partial x_i^\beta \partial x_i^\gamma \partial x_h^\tau}\bigg|_{\mathbf{x}=\bar{\mathbf{x}}} \\
&\quad + \frac{1}{2} \frac{\partial^4 U_h(r_{hi})}{\partial x_i^\alpha \partial x_i^\beta \partial x_i^\gamma \partial x_h^\tau}\bigg|_{\mathbf{x}=\bar{\mathbf{x}}} \\
q_{iihh}^{\alpha\beta\gamma\tau}(\bar{\mathbf{x}}) &= q_{ihih}^{\alpha\beta\gamma\tau}(\bar{\mathbf{x}}) = q_{hihi}^{\alpha\beta\gamma\tau}(\bar{\mathbf{x}}) \\
&= q_{hhii}^{\alpha\beta\gamma\tau}(\bar{\mathbf{x}}) = q_{ihhi}^{\alpha\beta\gamma\tau}(\bar{\mathbf{x}}) = q_{hiih}^{\alpha\beta\gamma\tau}(\bar{\mathbf{x}}) \\
&= \frac{1}{2} \frac{\partial^4 U_i(r_{ih})}{\partial x_i^\alpha \partial x_i^\beta \partial x_h^\gamma \partial x_h^\tau}\bigg|_{\mathbf{x}=\bar{\mathbf{x}}} \\
&\quad + \frac{1}{2} \frac{\partial^4 U_h(r_{hi})}{\partial x_i^\alpha \partial x_i^\beta \partial x_h^\gamma \partial x_h^\tau}\bigg|_{\mathbf{x}=\bar{\mathbf{x}}}
\end{aligned}
\tag{A.6}
$$

(iii) $i \neq j \neq k = h$ or $i \neq j \neq k = h$, $i, j, k, h = 1, 2, \ldots, N$ and $\alpha, \beta, \gamma, \tau = 1, 2, 3$.

$$
\begin{aligned}
q_{ijkh}^{\alpha\beta\gamma\tau}(\bar{\mathbf{x}}) &= q_{ijkk}^{\alpha\beta\gamma\tau}(\bar{\mathbf{x}}) = q_{jikk}^{\alpha\beta\gamma\tau}(\bar{\mathbf{x}}) \\
&= q_{kkij}^{\alpha\beta\gamma\tau}(\bar{\mathbf{x}}) = q_{kkji}^{\alpha\beta\gamma\tau}(\bar{\mathbf{x}}) \\
&= q_{kijk}^{\alpha\beta\gamma\tau}(\bar{\mathbf{x}}) = q_{kjik}^{\alpha\beta\gamma\tau}(\bar{\mathbf{x}}) \\
&= q_{ikjk}^{\alpha\beta\gamma\tau}(\bar{\mathbf{x}}) = q_{jkik}^{\alpha\beta\gamma\tau}(\bar{\mathbf{x}}) \\
&= q_{kikj}^{\alpha\beta\gamma\tau}(\bar{\mathbf{x}}) = q_{kjki}^{\alpha\beta\gamma\tau}(\bar{\mathbf{x}}) \\
&= q_{ikkj}^{\alpha\beta\gamma\tau}(\bar{\mathbf{x}}) = q_{jkki}^{\alpha\beta\gamma\tau}(\bar{\mathbf{x}}) = 0.
\end{aligned}
\tag{A.7}
$$

## Appendix 2: The proof of $g_{iii}^{\alpha\beta\gamma}(\bar{\mathbf{x}}) = 0$

**Proof** According to the definition of (11) and the expression (A.2), the term $g_{iii}^{\alpha\beta\gamma}(\bar{\mathbf{x}})$ is expressed as

$$
\begin{aligned}
g_{iii}^{\alpha\beta\gamma}(\bar{\mathbf{x}}) &= \frac{\partial^3 U(\mathbf{x})}{\partial x_i^\alpha \partial x_i^\beta \partial x_i^\gamma}\bigg|_{\mathbf{x}=\bar{\mathbf{x}}} \\
&= \frac{1}{2} \sum_{p=1, p \neq i}^{N_E} \frac{\partial^3 U_i(r_{ip})}{\partial x_i^\alpha \partial x_i^\beta \partial x_i^\gamma}\bigg|_{\mathbf{x}=\bar{\mathbf{x}}} \\
&\quad + \frac{1}{2} \sum_{l=1, l \neq i}^{N} \frac{\partial^3 U_l(r_{li})}{\partial x_i^\alpha \partial x_i^\beta \partial x_i^\gamma}\bigg|_{\mathbf{x}=\bar{\mathbf{x}}}.
\end{aligned}
\tag{A.8}
$$

where

$$
\begin{aligned}
\frac{\partial^3 U_i(r_{ip})}{\partial x_i^\alpha \partial x_i^\beta \partial x_i^\gamma}\bigg|_{\mathbf{x}=\bar{\mathbf{x}}} &= \bigg[ \frac{\partial^3 U_i(r_{ip})}{\partial (r_{ip})^3} \cdot \frac{\partial r_{ip}}{\partial x_i^\alpha} \cdot \frac{\partial r_{ip}}{\partial x_i^\beta} \cdot \frac{\partial r_{ip}}{\partial x_i^\gamma} \\
&\quad + \frac{\partial^2 U_i(r_{ip})}{\partial (r_{ip})^2} \cdot \bigg( \frac{\partial^2 r_{ip}}{\partial x_i^\beta \partial x_i^\gamma} \cdot \frac{\partial r_{ip}}{\partial x_i^\alpha} + \frac{\partial^2 r_{ip}}{\partial x_i^\alpha \partial x_i^\gamma} \cdot \frac{\partial r_{ip}}{\partial x_i^\beta} \\
&\quad + \frac{\partial^2 r_{ip}}{\partial x_i^\alpha \partial x_i^\beta} \cdot \frac{\partial r_{ip}}{\partial x_i^\gamma} \bigg) + \frac{\partial U_i(r_{ip})}{\partial r_{ip}} \cdot \frac{\partial^3 r_{ip}}{\partial x_i^\alpha \partial x_i^\beta \partial x_i^\gamma} \bigg]\bigg|_{\mathbf{x}=\bar{\mathbf{x}}}, \\
\frac{\partial^3 U_l(r_{li})}{\partial x_i^\alpha \partial x_i^\beta \partial x_i^\gamma}\bigg|_{\mathbf{x}=\bar{\mathbf{x}}} &= \bigg[ \frac{\partial^3 U_l(r_{li})}{\partial (r_{li})^3} \cdot \frac{\partial r_{li}}{\partial x_i^\alpha} \cdot \frac{\partial r_{li}}{\partial x_i^\beta} \cdot \frac{\partial r_{li}}{\partial x_i^\gamma} \\
&\quad + \frac{\partial^2 U_l(r_{li})}{\partial (r_{li})^2} \cdot \bigg( \frac{\partial^2 r_{li}}{\partial x_i^\beta \partial x_i^\gamma} \cdot \frac{\partial r_{li}}{\partial x_i^\alpha} + \frac{\partial^2 r_{li}}{\partial x_i^\alpha \partial x_i^\gamma} \cdot \frac{\partial r_{li}}{\partial x_i^\beta} \\
&\quad + \frac{\partial^2 r_{li}}{\partial x_i^\alpha \partial x_i^\beta} \cdot \frac{\partial r_{li}}{\partial x_i^\gamma} \bigg) + \frac{\partial U_l(r_{li})}{\partial r_{li}} \cdot \frac{\partial^3 r_{li}}{\partial x_i^\alpha \partial x_i^\beta \partial x_i^\gamma} \bigg]\bigg|_{\mathbf{x}=\bar{\mathbf{x}}}.
\end{aligned}
\tag{A.9}
$$

Since the RVE we considered is in transient equilibrium state, each atom is in equilibrium of forces at any direction. Thus, there exist paired centrosymmetric atoms around each atom. As is shown in Fig. 8, the $p$th and $q$th atoms are centrosymmetric with respect to the $i$th atom. And $\bar{r}_{ip} = \bar{r}_{iq}$, where $\bar{r}_{ip}$ is the distance between the $i$th and $p$th atoms, and $\bar{r}_{qi}^\alpha$ is the distance between the $i$th and $q$th atoms. It is easy to obtain that

$$
\begin{aligned}
\frac{\partial r_{ip}}{\partial x_i^\alpha}\bigg|_{\mathbf{x}=\bar{\mathbf{x}}} &= -\frac{\partial r_{iq}}{\partial x_i^\alpha}\bigg|_{\mathbf{x}=\bar{\mathbf{x}}}, \\
\frac{\partial r_{ip}}{\partial x_i^\alpha \partial x_i^\beta}\bigg|_{\mathbf{x}=\bar{\mathbf{x}}} &= \frac{\partial r_{iq}}{\partial x_i^\alpha \partial x_i^\beta}\bigg|_{\mathbf{x}=\bar{\mathbf{x}}}, \\
\frac{\partial^3 r_{ip}}{\partial x_i^\alpha \partial x_i^\beta \partial x_i^\gamma}\bigg|_{\mathbf{x}=\bar{\mathbf{x}}} &= -\frac{\partial^3 r_{iq}}{\partial x_i^\alpha \partial x_i^\beta \partial x_i^\gamma}\bigg|_{\mathbf{x}=\bar{\mathbf{x}}}, \quad \forall \alpha, \beta, \gamma = 1, 2, 3.
\end{aligned}
$$

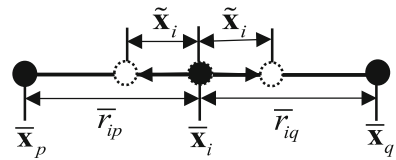

**Fig. 8** The equilibrium state of $i$th oscillator at $\alpha(\alpha = 1, 2, 3)$ direction

$$\frac{\partial U_i(r_{ip})}{\partial r_{ip}}\bigg|_{\mathbf{x}=\bar{\mathbf{x}}} = \frac{\partial U_i(r_{iq})}{\partial r_{iq}}\bigg|_{\mathbf{x}=\bar{\mathbf{x}}},$$

$$\frac{\partial^2 U_i(r_{ip})}{\partial (r_{ip})^2}\bigg|_{\mathbf{x}=\bar{\mathbf{x}}} = \frac{\partial^2 U_i(r_{iq})}{\partial (r_{iq})^2}\bigg|_{\mathbf{x}=\bar{\mathbf{x}}},$$

$$\frac{\partial^3 U_i(r_{ip})}{\partial (r_{ip})^3}\bigg|_{\mathbf{x}=\bar{\mathbf{x}}} = \frac{\partial^3 U_i(r_{iq})}{\partial (r_{iq})^3}\bigg|_{\mathbf{x}=\bar{\mathbf{x}}}, \quad \forall i = 1, 2, \ldots, N.$$

$$\text{(A.10)}$$

Substituting the relations (A.10) into (A.9), it yields that

$$\frac{\partial^3 U_i(r_{ip})}{\partial x_i^\alpha \partial x_i^\beta \partial x_i^\gamma}\bigg|_{\mathbf{x}=\bar{\mathbf{x}}} + \frac{\partial^3 U_i(r_{iq})}{\partial x_i^\alpha \partial x_i^\beta \partial x_i^\gamma}\bigg|_{\mathbf{x}=\bar{\mathbf{x}}} = 0,$$

$$\frac{\partial^3 U_p(r_{pi})}{\partial x_i^\alpha \partial x_i^\beta \partial x_i^\gamma}\bigg|_{\mathbf{x}=\bar{\mathbf{x}}} + \frac{\partial^3 U_q(r_{qi})}{\partial x_i^\alpha \partial x_i^\beta \partial x_i^\gamma}\bigg|_{\mathbf{x}=\bar{\mathbf{x}}} = 0.$$

$$\text{(A.11)}$$

Finally, combining (A.8)–(A.11), the proof $g_{iii}^{\alpha\beta\gamma}(\bar{\mathbf{x}}) = 0$ is completed. □

# References

1. Born M, Huang K (1955) Dynamical theory of crystal lattices. Oxford and Clarendon Press, London
2. Kittel C (2005) Introduction to solid state physics, 8th edn. Wiley, New York
3. Wallace DC (2002) Statistical physics of crystals and liquids: a guide to highly accurate equations of state. World Scientific, Singapore
4. Krishnan RS, Srinivasan R, Devanarayanan S (2013) Thermal expansion of crystals: international series in the science of the solid state, vol 12. Pergamom Press Ltd, Oxford
5. Barron THK, White GK (2012) Heat capacity and thermal expansion at low temperatures. Springer, New York
6. van de Walle A, Ceder G (2002) The effect of lattice vibrations on substitutional alloy thermodynamics. Rev Mod Phys 74:11–45
7. Fultz B (2010) Vibrational thermodynamics of materials. Prog Mater Sci 55(4):247–352
8. Moruzzi VL, Janak JF, Schwarz K (1988) Calculated thermal properties of metals. Phys Rev B 37:790–799
9. Gan CK, Soh JR, Liu Y (2015) Large anharmonic effect and thermal expansion anisotropy of metal chalcogenides: the case of antimony sulfide. Phys Rev B 92:235202
10. Glensk A, Grabowski B, Hickel T, Neugebauer J (2015) Understanding anharmonicity in fcc materials: from its origin to ab initio strategies beyond the quasiharmonic approximation. Phys Rev Lett 114:195901
11. Grabowski B, Ismer L, Hickel T, Neugebauer J (2009) Ab initio up to the melting point: anharmonicity and vacancies in aluminum. Phys Rev B 79:134106
12. Hellman O, Abrikosov IA, Simak SI (2011) Lattice dynamics of anharmonic solids from first principles. Phys Rev B 84:180301
13. Monserrat B, Drummond ND, Needs RJ (2013) Anharmonic vibrational properties in periodic systems: energy, electron–phonon coupling, and stress. Phys Rev B 87:144302
14. Bansal D, Aref A, Dargush G, Delaire O (2016) Modeling non-harmonic behavior of materials from experimental inelastic neutron scattering and thermal expansion measurements. J Phys Condens Matter 28(38):385201
15. Narasimhan S, De Gironcoli S (2002) Ab initio calculation of the thermal properties of Cu: performance of the LDA and GGA. Phys Rev B 65(6):064302
16. Mounet N, Marzari N (2005) First-principles determination of the structural, vibrational and thermodynamic properties of diamond, graphite, and derivatives. Phys Rev B 71(20):205214
17. Yun Y, Legut D, Oppeneer PM (2012) Phonon spectrum, thermal expansion and heat capacity of $UO_2$ from first-principles. J Nucl Mater 426(1):109–114
18. Allen PB (2015) Anharmonic phonon quasiparticle theory of zero-point and thermal shifts in insulators: heat capacity, bulk modulus, and thermal expansion. Phys Rev B 92:064106
19. Liu ZJ, Song T, Sun XW, Ma Q, Wang T, Guo Y (2017) Thermal expansion, heat capacity and Grneisen parameter of iridium phosphide $Ir_2P$ from quasi-harmonic Debye mode. Solid State Commun 253:19–23
20. Li H, Cui J, Li B (2015) A thermo-mechanical coupling atom–continuum couple model and it algorithm. Appl Math Mech Engl 36(4):343–351
21. Xiang M, Cui J, Li B, Tian X (2012) Atom-continuum coupled model for thermo-mechanical behavior of materials in micro–nano scales. Sci China Phys Mech 55(6):1125–1137
22. Li B, Cui J, Tian X, Xingang Y, Xiang M (2014) The calculation of mechanical behavior for metallic devices at nano-scale based on atomic–continuum coupled model. Comput Mater Sci 94(11):73–84
23. To AC, Liu WK, Kopacz A (2008) A finite temperature continuum theory based on interatomic potential in crystalline solids. Comput Mech 42(4):531–541
24. Gallavotti G (2013) Statistical mechanics. Springer, New York
25. Arnol'd Vladimir Igorevich (2013) Mathematical methods of classical mechanics. Springer, New York
26. Kovacic I, Brennan MJ (2011) The Duffing equation: nonlinear oscillators and their behaviour. Wiley, New York
27. Liu S, Zuntao F, Liu S, Zhao Q (2001) Jacobi elliptic function expansion method and periodic wave solutions of nonlinear wave equations. Phys Lett A 289(1):69–74
28. Parkes EJ, Duffy BR, Abbott PC (2002) The Jacobi elliptic-function method for finding periodic-wave solutions to nonlinear evolution equations. Phys Lett A 295(56):280–286
29. Mishin Y, Mehl MJ, Papaconstantopoulos DA, Voter AF, Kress JD (2001) Structural stability and lattice defects in copper: Ab initio, tight-binding, and embedded-atom calculations. Phys Rev B 63:224106
30. Lee BJ, Shim JH, Baskes MI (2003) Semiempirical atomic potentials for the fcc metals Cu, Ag, Au, Ni, Pd, Pt, Al, and Pb based on first and second nearest-neighbor modified embedded atom method. Phys Rev B 68:144112
31. Foiles SM, Baskes MI, Daw MS (1986) Embedded-atom-method functions for the fcc metals Cu, Ag, Au, Ni, Pd, Pt, and their alloys. Phys Rev B 33:7983–7991
32. Williams PL, Mishin Y, Hamilton JC (2006) An embedded-atom potential for the Cu–Ag system. Model Simul Mater Sci 14(5):817
33. Gray DE (1972) American institute of physics handbook. McGraw-Hill, New York
34. Touloukian YS, Kirby RK, Taylor RE, Desai PD (1975) Thermal expansion: metallic elements and alloys, vol 12. Springer, New York